\documentclass[twocolumn,english,aps,runinaddress,preprintnumbers,floatfix,showpacs,prb]{revtex4}
\usepackage[T1]{fontenc}
\usepackage[latin9]{inputenc}
\usepackage{color}
\usepackage{bm}
\usepackage{amsmath}
\usepackage{amssymb}
\usepackage{graphicx}
\usepackage{esint}
\PassOptionsToPackage{normalem}{ulem}
\usepackage{ulem}

\makeatletter
\@ifundefined{textcolor}{}
{%
 \definecolor{BLACK}{gray}{0}
 \definecolor{WHITE}{gray}{1}
 \definecolor{RED}{rgb}{1,0,0}
 \definecolor{GREEN}{rgb}{0,1,0}
 \definecolor{BLUE}{rgb}{0,0,1}
 \definecolor{CYAN}{cmyk}{1,0,0,0}
 \definecolor{MAGENTA}{cmyk}{0,1,0,0}
 \definecolor{YELLOW}{cmyk}{0,0,1,0}
 }

\@ifundefined{definecolor}{\usepackage{color}}{}
\@ifundefined{definecolor}{\usepackage{color}}{}

\definecolor{darkblue}{rgb}{0,0,.5}
\definecolor{darkred}{rgb}{.5,0,0}
\definecolor{darkyellow}{cmyk}{0,0,1,0.4}

\newcommand{\XSTAR}{{x^{\star}}}
\usepackage{babel}
\usepackage{hyperref}

 \renewcommand{\cite}{\onlinecite}

\makeatother

\usepackage{babel}
\begin{document}

\author{M. Greger}

\author{M. Kollar}

\author{D. Vollhardt}

\affiliation{Theoretical Physics III, Center for Electronic Correlations and Magnetism,
Institute of Physics, University of Augsburg, 86135 Augsburg, Germany}

\begin{abstract}
We analyze the sharpness of crossing ({}``isosbestic'') points of
a family of curves which are observed in many quantities described
by a function $f(x,p)$, where $x$ is a variable (e.g., the frequency)
and $p$ a parameter (e.g., the temperature). We show that if a narrow
crossing region is observed near $x_{*}$ for a range of parameters
$p$, then $f(x,p)$ can be approximated by a perturbative expression
in $p$ for a wide range of $x$. This allows us, e.g.,  to extract
the temperature dependence of several experimentally obtained quantities,
such as the Raman response of HgBa$_{2}$CuO$_{4+\delta}$, photoemission
spectra of thin VO$_{2}$ films, and the reflectivity of CaCu$_{3}$Ti$_{4}$O$_{12}$,
all of which exhibit narrow crossing regions near certain frequencies.
We also explain the sharpness of isosbestic points in the optical
conductivity of the Falicov-Kimball model and the spectral function
of the Hubbard model. 
\end{abstract}

\title{Isosbestic points: How a narrow crossing region of curves
determines their leading parameter dependence}

\maketitle

\section{\label{sec:intro}Introduction}

As early as 1937 it was noticed {[}\cite{Scheibe1937}{]} that the
curves of the extinction coefficient $\alpha(\omega,c_{1})$ of a
mixture of two liquid solutions 
intersect exactly at one frequency $\omega^{\star}$ when plotted
as a function of $\omega$ for different concentrations $c_{1}$ of
one of the components. 
At this particular frequency the extinction coefficient therefore
no longer depends on the parameter $c_{1}$. This point of intersection
of the family of curves $\alpha(\omega,c_{1})$ is referred to as
\emph{isosbestic point} {[}\cite{cohen1962ip,morrey1963isosbestic,isosbestic}{]}.
Today isosbestic points 
of the extinction coefficient of two compounds 
play an important role in spectrophotometry {[}\cite{gore2000spectrophotometry}{]}.

It is a common practice in the natural sciences to illustrate the
functional dependence of a quantity $f$ on a variable $x$ and a
parameter $p$ by plotting $f(x,p)$ vs. $x$ for several values of
$p$, say $p_{1},\dots p_{n}$. This leads to a family of $n$ curves
$f(x,p_{i})$, $i=1,\dots n$. The curves will usually intersect.
If the points of intersection are confined to a narrow region of $x$
values, or if they even coincide at a single point $x^{\star}$ as
in the case of the absorbance $\alpha(\omega,c_{1})$, it leads to
a conspicuous feature whose origin calls for an explanation. Well-defined
crossing points of curves are now often called {}``isosbestic points''
even if they do not fall onto a single point {[}\cite{PhysRevB.43.7942,PhysRevB.64.233114}{]}.
We will use this term also in the present paper. We stress that the
characteristic $p$-independence of $f(x,p)$ at $x^{\star}$ is these
general cases only fulfilled in a fairly small $p$-interval, i.e.,
it is a {}``local'' phenomenon.

For example, an unexpected sharp crossing point at $T^{\star}=160$mK
was recognized in the family of curves of the specific heat $c_{V}(T)$
of normal-liquid $^{3}$He when plotted as a function of temperature
for different molar volumes {[}\cite{PhysRevB.27.2747}{]}. Rather
sharp crossing points are also observed {[}\cite{vollhardt1997ccp}{]}
in the specific heat curves $c(T,X)$ 
of many heavy-fermion compounds, with $X$ as a second thermodynamic
variable, e.g., when measured at different pressures ($X=P$) as in
CeAl$_{3}$ {[}\cite{PhysRevLett.56.390}{]}, or for different magnetic
fields ($X=H$) as in CeCu$_{6-x}$Al$_{x}$ {[}\cite{Schlager1993}{]},
RuSr$_{2}$Gd$_{1.5}$Ce$_{0.5}$Cu$_{2}$O$_{10-\delta}$ {[}\cite{belevtsev2009characteristic}{]},
Mn$_{1-x}$Fe$_{x}$Si {[}\cite{PhysRevB.82.064404}{]}, Mn$_{1-x}$Co$_{x}$Si
{[}\cite{PhysRevB.82.064404}{]}, Cu$_{2}$OSeO$_{3}$ {[}\cite{janoschek2012fluctuation}{]},
and MnSi {[}\cite{PhysRevLett.108.237204}{]}. 

Crossing points of specific heat curves have been found in theoretical
investigations of lattice models for correlated electrons such as
the one-band Hubbard model {[}\cite{vollhardt1997ccp,Seiler1986,PhysRevB.48.7167,chandra1999nuc}{]}.
At half-filling the curves obtained by plotting $c(T,U)$ vs.\ $T$
for different values of the local Coulomb repulsion $U$ always cross
at two temperatures, irrespective of the type of the lattice, as seen
in the case of nearest-neighbor hopping in $d=1$ {[}\cite{shiba1972tpo,PTP.48.2171,juttner1998hcf}{]},
$d=2$ {[}\cite{PhysRevB.55.12918,PhysRevB.63.125116,PhysRevB.65.153109,PhysRevB.72.085123}{]},
and $d=\infty$ {[}\cite{PhysRevB.48.7167}{]}, as well as for long-range
hopping in $d=1$ {[}\cite{PhysRevLett.68.244}{]}. In particular,
the crossing point at high temperatures was found to occur at a nearly
universal value $c^{\star}/k_{B}\simeq0.34$ {[}\cite{chandra1999nuc,High-T_crossing_point}{]}.
Subsequently crossing points were noticed also in the charge susceptibility
of the half-filled extended Hubbard model in $d=1$ {[}\cite{PhysRevB.76.155121}{]}
and the thermal conductance of strongly correlated quantum dots {[}\cite{PhysRevB.81.235127}{]}.
The heavy fermion compound YbRh$_{2}$Si$_{2}$ exhibits analogous isosbestic
points in its magnetic susceptibility $\chi(H,T)$ {[}\cite{PSSB:PSSB201200771}{]}.

The crossing points discussed above all occur when the temperature
dependence of some physical quantity such as the specific heat is
plotted for different values of a parameter, e.g., pressure, magnetic
field, or interaction strength. But crossing points have also been
observed in experimental data and theoretical results for the frequency
dependence of various response functions, e.g., (i) in the optical
conductivity $\sigma(\omega,X)$ of electron doped Nd$_{2-\delta}$Ce$_{\delta}$CuO$_{4}$,
with $X=\delta$ as the doping level {[}\cite{PhysRevB.43.7942}{]},
and other high-T$_{c}$ materials {[}\cite{RevModPhys.70.1039}{]}
as well as in the pyrochlore-type molybdate family R$_{2}$Mo$_{2}$O$_{7}$
(R=Nd, Sm, Eu, Gd) {[}\cite{PhysRevB.73.125122}{]}, where $X$ now
denotes the ionic radius $r$ associated with the different elements
R; (ii) in the Raman response function $\chi(\omega,T)$ of the high-T$_{c}$
material HgBa$_{2}$CuO$_{4+\delta}$ plotted as a function of the
Raman shift $\omega$ for different temperatures {[}\cite{guyard2008breakpoint}{]};
and (iii) in the dielectric function $\epsilon(\omega,T)$ of the
colossal magnetoresistance material LaMnO$_{3}$ {[}\cite{PhysRevLett.93.147204}{]}
and the reflectivity $R(\omega,T)$ of the colossal dielectric constant
material CaCu$_{3}$Ti$_{4}$O$_{12}$ {[}\cite{kant2008bdr}{]} when
plotted as a function of the photon energy $\omega$ for different
values of $T$.

Sharp crossing points in the Raman response $\chi(\omega,T)$ have
been found in theoretical investigations based on the Hubbard model
{[}\cite{PhysRevB.64.233114,PhysRevB.67.155102}{]} and the Falicov-Kimball
model {[}\cite{RevModPhys.75.1333,PhysRevLett.93.137402}{]}. The
spectral function $A(\omega,U)$ of the Hubbard model computed within
dynamical mean-field theory (DMFT) also shows a sharp crossing point
{[}\cite{eckstein2007ips}{]}.

Whenever isosbestic points occur in a function $f(x,p)$ two separate
questions arise {[}\cite{vollhardt1997ccp}{]}:
\begin{enumerate}
\item Why do the curves cross at all, and
\item why is the crossing feature confined to a narrow region despite significant
changes in the parameter $p$ {[}\cite{Linear-scale}{]}?
\end{enumerate}
As discussed by one of us {[}\onlinecite{vollhardt1997ccp}{]} the
existence of crossing points in specific heat curves can be explained
by a sum rule for the change of the entropy $S(T,X)$ with respect
to $X$ in the limit of $T\rightarrow\infty$. Furthermore, their
sharpness can be linked to the smallness of the susceptibilities $\chi^{(n)}(T,X)=\partial^{n}\xi/\partial X^{n}$
of strongly correlated systems in an expansion of the specific heat
$c(T,X)$ with respect to $X$, where $\xi(T,X)$ is the conjugate
variable to $X$ {[}\onlinecite{VINV}{]}.

   In this paper we concentrate on the second question.  In
   Sec.~\ref{sec:Quantitative-formulation} we discuss the properties
   of isosbestic points and derive the leading parameter dependence
   for a function $f(x,p)$ with an approximate isosbestic point. This yields
   a generalization of the two-fluid model {[}\cite{Scheibe1937}{]}
   and allows us to extract the leading parameter dependence for
   quantities which exhibit such a narrow crossing region.  In
   Sec.~\ref{sec:Applications} we then extract the temperature
   dependence of several experimentally obtained quantities, such as
   the Raman response $\chi\left(\omega,T\right)$ of the cuprate
   compound HgBa$_{2}$CuO$_{4+\delta}$, in photoemission spectra
   $I\left(\omega,T\right)$ of VO$_{2}$ thin films, and in the
   reflectivity $R\left(\omega,T\right)$ of
   CaCu$_{3}$Ti$_{4}$O$_{12}$. We also explain the sharpness of
   isosbestic points in the optical conductivity of the
   Falicov-Kimball model {[}\cite{RevModPhys.75.1333}{]}
   (Sec.~\ref{sec:FK}) and the spectral function
   $A\left(\omega,U\right)$ of the Hubbard model
   (Sec.~\ref{sec:Hubbard}), both within DMFT
   {[}\cite{PhysRevLett.62.324,PhysRevB.45.6479,vollhardt1993correlated,pruschke1995anomalous,RevModPhys.68.13,kotliar2004strongly}{]}.
   A conclusion follows in Sec.~\ref{sec:Conclusion}.

\section{\label{sec:Quantitative-formulation}Properties of exact and approximate
isosbestic points}

\subsection{Existence of crossing points}

A family of non-monotonic curves, obtained by plotting a quantity
$f(x,p)$ as a function of $x$ for different values of a parameter
$p$, will in general intersect. The crossing points are located along
a curve $x^{\star}(p)$ defined by {[}\cite{vollhardt1997ccp,eckstein2007ips}{]}
\begin{equation}
\frac{\partial f(x,p)}{\partial p}\bigg|_{x^{\star}(p)}=0\,.\label{crossing}
\end{equation}
 If these points of intersection are confined to a narrow region (in
which case we refer to them as isosbestic points), the value $x^{\star}(p)$
depends only weakly on $p$. If the curves intersect at a single point,
$x^{\star}$ does not depend on $p$ at all.

Eq.~\eqref{crossing} immediately explains why the absorbance of
a mixture of two liquid solutions with individual concentrations $c_{1},c_{2}$,
whose total concentration is constant ($c_{1}+c_{2}=c$), has a sharp
isosbestic point {[}\cite{cohen1962ip}{]}. Namely, in this case the
absorbance $\alpha(\omega,c_{1})$ depends only linearly on the concentration
$c_{1}$, i.e., has the special form
\begin{equation}
\alpha(\omega,c_{1})=c_{1}\alpha_{1}(\omega)+(c-c_{1})\alpha_{2}(\omega).\label{absorbance}
\end{equation}
 If $\alpha_{1}$ and $\alpha_{2}$ coincide at some frequency $\omega^{\star}$,
i.e., $\alpha_{1}(\omega^{\star})=\alpha_{2}(\omega^{\star})$, then
\begin{equation}
\frac{\partial\alpha(\omega,c_{1})}{\partial c_{1}}\bigg|_{\omega^{\star}}=0\,.
\end{equation}
 This implies that for all concentrations $c_{1}$ the absorbance
curves intersect exactly at one frequency $\omega^{\star}$, or the
equivalent wavelength. Quite generally, whenever a system is a superposition
of two components, where the sum of the densities is conserved, isosbestic
points in the curves plotted for different densities are bound to
occur {[}\cite{eckstein2007ips,Two-fluid}{]}. Indeed, this argument
is used by Uchida \emph{et al.} {[}\cite{PhysRevB.43.7942}{]} and
{K\'{e}zsm\'{a}rki} \emph{et al.} {[}\cite{PhysRevB.73.125122}{]} to explain
the well-defined isosbestic points in the optical conductivity $\sigma(\omega,p)$
of numerous correlated electron materials, where $p$ is some control
parameter, e.g., the temperature or the bandwidth. Starting from the
sum rule for the optical conductivity, and assuming (i) that the spectral
weight is only redistributed between electrons with two different
energy scales, and that (ii) the optical spectrum can be decomposed
and linearly interpolated by two terms as a function of $p$, they
arrive at an expression for the optical conductivity given by $\sigma(\omega,p)=p\sigma_{1}(\omega)+(1-p)\sigma_{2}(\omega)$,
which is identical to the result for the absorbance in Eq.~\eqref{absorbance}
and hence leads to an isosbestic point at some particular frequency
$\omega^{\star}$ {[}\cite{private-communication}{]}.

Before we continue with our discussion we need to introduce a suitable
nomenclature. Based on the behavior of the curve of crossing points
$x^{\star}(p)$ we identify three different cases: (i) A globally
\textit{exact} isosbestic point, which is characterized by $x^{\star}(p)=const$,
corresponding to a complete $p$-independence of $f(x^{\star},p)$;
(ii) a locally exact isosbestic point, which arises when $x^{\star}(p)$
exhibits a local extremum or a higher order stationary point around
some value $p=p_{0}$, corresponding to a locally $p$-independent
$f(x^{\star}(p),p)$ around $p=p_{0}$; (iii) an approximate isosbestic
point, which corresponds to a weak $p$-dependence of $x^{\star}(p)$.

We note that locally exact isosbestic points can be further classified
by the order of the stationary point of $x^{\star}(p)$ at $p_{0}$.
If
\begin{equation}
\frac{\partial^{k}x^{\star}(p)}{\partial p^{k}}\bigg|_{p_{0}}=0\label{eq:k_th_order_cond}
\end{equation}
for $k\leq n$, the isosbestic point is called an {}``isosbestic
point of $n$th order''.

\subsection{\label{sub:Sharpness-of-crossing}Sharpness of isosbestic points}

To understand the general origin of sharp isosbestic points in a function
$f(x,p)$, it is important to note that isosbestic behavior is usually
observed only in a certain parameter range around some particular
value $p_{0}$ and can be expected to break down away from $p_{0}$.
Accordingly, the weak $p$-dependence of $\XSTAR(p)$ required for
an approximate isosbestic point will generally be a local phenomenon,
i.e., it applies only for a finite $p$-interval around $p_{0}$.
This observation of the local nature of isosbestic points motivates
the following expansion around $p_{0}$
\begin{equation}
f\left(x,p\right)=f\left(x,p_{0}\right)+(p-p_{0})\, F_{1}(x,p_{0})+\mathcal{O}\left[(p-p_{0})^{2}\right],\label{eq:gen_expand}
\end{equation}
where $F_{n}(x,p)=\partial^{n}f(x,p)/\partial p^{n}$. One has
$F_{1}(\XSTAR(p_{0}),p_{0})$ $=$ $0$,
by virtue of the definition of $\XSTAR(p)$ {[}Eq.~\eqref{crossing}{]}.
As a consequence, the linear approximation of $f(x,p)$
\begin{equation}
f_{\text{approx}}(x,p)=f(x,p_{0})+(p-p_{0})\, F_{1}(x,p_{0})\,,\label{eq:f_approx}
\end{equation}
will exhibit an exact isosbestic point at $x=\XSTAR(p_{0})$ for all
values of $p$. Since
\begin{equation}
f\left(x,p\right)=f_{\text{approx}}(x,p)+\mathcal{O}\left[(p-p_{0})^{2}\right]\label{eq:gen_expand_interpret}
\end{equation}
 the second term is seen to be responsible for deviations from the
exact isosbestic point described by $f_{\text{approx}}(x,p)$. This
result can be put to a test by checking the validity of the following
approximation:
\begin{eqnarray}
\widetilde{f}\left(x,p\right) & \equiv & f(x,p)-(p-p_{0})F_{1}(x,p_{0})\label{eq:Delta_f}\\
 & = & f(x,p_{0})+\mathcal{O}\left[(p-p_{0})^{2}\right]\label{eq:Delta_f2}\\
 & \simeq & f(x,p_{0})\,.\label{eq:Delta_f3}
\end{eqnarray}
The quantity $\widetilde{f}\left(x,p\right)$ represents the reference
curve $f(x,p_{0})$ plus all terms which are responsible for the deviations
from the exact isosbestic case. The verification of its approximate
parameter independence, i.e., the verification of the approximation
$\widetilde{f}\left(x,p\right)\simeq f(x,p_{0})$ allows one to quantify
the importance of the subleading terms $\mathcal{O}\left[(p-p_{0})^{2}\right]$
and therefore the deviations from the exact isosbestic case.

In general, the value of $p_{0}$ should correspond to the position
of the weakest $p$-dependence of $\XSTAR(p)$ and, without any further
information on the system, is generally hard to determine unambiguously.
Of particular interest however are the cases where $p_{0}$ corresponds
to an extremum or some higher-order stationary point of $\XSTAR(p)$.
These cases lead to locally exact isosbestic points where $p_{0}$
is determined by $\XSTAR^{\prime}(p_{0})=0$. The new variable $(p-p_{0})$
is then a small parameter of the system, e.g., an internal frequency
scale if $p$ is a frequency. From the implicit Eq.~\eqref{crossing}
one obtains
\begin{align}
\frac{d}{dp}\XSTAR(p) & =-\left.\frac{{\partial F_{1}(x,p)}/{\partial p}}{{\partial F_{1}(x,p)}/{\partial x}}\right|_{x=\XSTAR(p)}\,,
\end{align}
which means that for $\XSTAR^{\prime}(p_{0})=0$ the subleading terms
in Eq.~\ref{eq:gen_expand_interpret} are at least of order $(p-p_{0})^{3}$,
thus corresponding to particularly weak deviations from the exact
isosbestic point in $f_{\text{approx}}(x,p)$ in a vicinity of $p_{0}$.
In practice isosbestic points corresponding to an extremum of $\XSTAR(p)$
are rather common, as will be discussed in the following sections.
A more mathematical treatment regarding the behavior around such stationary
points of $\XSTAR(p)$ is provided in Appendix \ref{sec:Size-of-the}.

In the rest of the paper we will use these results to show that isosbestic
points are indeed connected to a leading parameter dependence as described
by Eq.~\eqref{eq:f_approx}.

Finally we emphasize that it is  not possible to quantify the sharpness of isosbestic points
beyond Eqs.~(\ref{eq:Delta_f}-\ref{eq:Delta_f3}), since the \textit{perceived} sharpness
always depends on the magnification of the crossing region, 
which is not an intrinsic  property of $f(x,p)$
itself (see Appendix \ref{sec:SharpnessS}).

\section{\label{sec:Applications}Analysis of experimentally observed isosbestic
points}

The generality of Eq.~\eqref{eq:gen_expand_interpret} allows us
to investigate the sharpness of a whole class of isosbestic points
in entirely unrelated systems within a common framework. An existing
set of theoretical or experimental data of a quantity $Q\left(x,p_{i}\right)$
obtained for a number of different parameters, $i$ $=$ $1,2,\ldots n$
can be analyzed in the vicinity of an isosbestic point $x^{\star}(p)$
by comparing it to
\begin{eqnarray}
Q_{\text{approx}}\left(x,p\right) & = & Q(x,p_{0})+(p-p_{0})Q_{1}(x,p_{0}).\label{eq:gen_expand_experiment}
\end{eqnarray}
 Here the function $Q_{1}\left(x,p_{0}\right)$ is approximated as
\begin{equation}
Q_{1}(x,p_{0})\simeq\frac{Q(x,p_{\alpha})-Q(x,p_{\beta})}{(p_{\alpha}-p_{0})-(p_{\beta}-p_{0})},\label{eq:approx_Q1}
\end{equation}
with $p_{\alpha},p_{\beta}\in[p_{1}\ldots p_{n}]$ sufficiently close
to $p_{0}$ to obtain a proper approximation of $Q_{1}(x,p_{0})$.
As before the applicability of Eq.~\eqref{eq:gen_expand_experiment}
is then tested by verifying the approximate $p$ independence of
\begin{eqnarray}
\widetilde{Q}(x,p_{i}) & = & Q(x,p_{i})-(p-p_{0})Q_{1}(x,p_{0})\label{eq:Delta_Q}\\
 & \simeq & Q(x,p_{0})\,.
\end{eqnarray}

\begin{figure}[t]
\centering{}\includegraphics[scale=0.95]{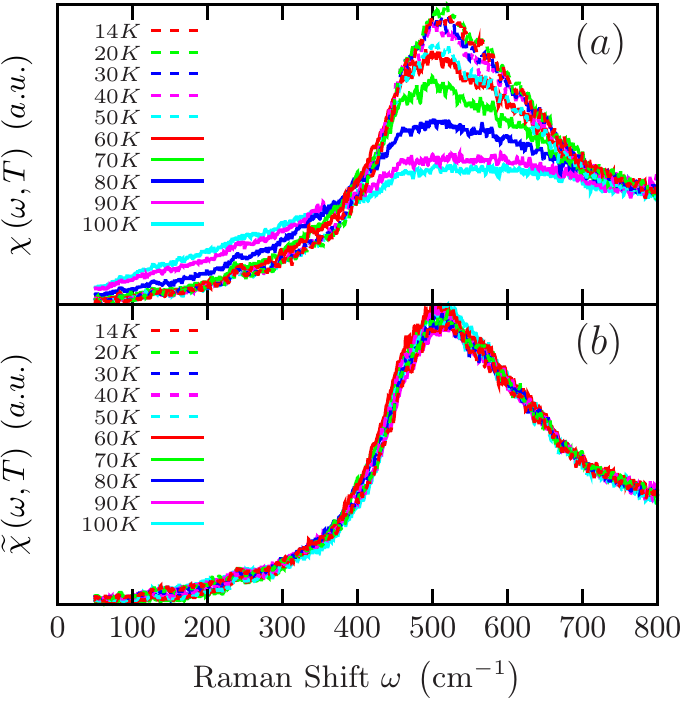}\caption{\label{fig:Fig6}(a)
  Antinodal Raman response $\chi\left(\omega,T\right)$
of HgBa$_{2}$CuO$_{4+\delta}$ {[}\cite{guyard2008breakpoint}{]} at
optimal doping and (b) the
function $\widetilde{\chi}\left(\omega,T\right)$. The weak temperature
dependence of $\widetilde{\chi}\left(\omega,T\right)$ explains the
sharpness of the isosbestic point and also confirms the validity of
ansatz \eqref{eq:raman_ch_exp} within the available data.}
\end{figure}

\subsection{\label{sub:Raman-response-of}Raman response of HgBa$_{2}$CuO$_{4+\delta}$}

We now apply the scheme to the isosbestic point observed in measurements
{[}\cite{guyard2008breakpoint}{]} of the antinodal ($\bm{B}_{1g}$)
Raman response $\chi\left(\omega,T\right)$ of the cuprate compound
HgBa$_{2}$CuO$_{4+\delta}$ for various temperatures at optimal doping
($\delta$ $=$ $0.16$), which shows a distinct isosbestic point
for up to $\simeq100K$. Proceeding as described above we make the
following Sommerfeld-type ansatz for the low temperature dependence
of $\chi\left(\omega,T\right)$:
\begin{equation}
\chi\left(\omega,T\right)=\chi\left(\omega,0\right)+T^{2}\chi_{2}(\omega)+\mathcal{O}[T^{3}],\label{eq:raman_ch_exp}
\end{equation}
 where
\begin{equation}
\chi_{2}(\omega)\simeq\frac{\chi(\omega,T_{1})-\chi(\omega,T_{2})}{T_{1}^{2}-T_{2}^{2}},\label{eq:rm_fun1}
\end{equation}
 as in Eq.~\eqref{eq:approx_Q1}. Again, the validity of Eq.~\eqref{eq:raman_ch_exp}
can be tested via the $T$-independence of the quantity
\begin{eqnarray}
\widetilde{\chi}\left(\omega,T\right) & = & \chi\left(\omega,T\right)-T^{2}\chi_{2}(\omega)\nonumber \\
 & \simeq & \chi\left(\omega,T=0\right).\label{eq:raman_delta}
\end{eqnarray}
 This is done in Fig.~\ref{fig:Fig6}, which shows good agreement
with Eq.~\eqref{eq:raman_delta}. To determine $\chi_{2}(\omega)$
we chose $T_{1}=14K$ and $T_{2}=90K$, which is by no means a unique
combination {[}\cite{CommentCuprate}{]}. \sout{}

It is interesting to note that the isosbestic point, and thus the
quadratic $T$ dependence in Eq.~\eqref{eq:raman_ch_exp}, are specific
features of the \textit{optimally doped} system and are absent for
other dopings {[}\cite{guyard2008breakpoint}{]} (at least for the
provided temperatures). For the case of optimal doping, we have explained
the sharp isosbestic point in $\chi(\omega,T)$ by its essentially
quadratic temperature dependence.

\begin{figure}[t]

\centering{}\includegraphics[scale=0.55]{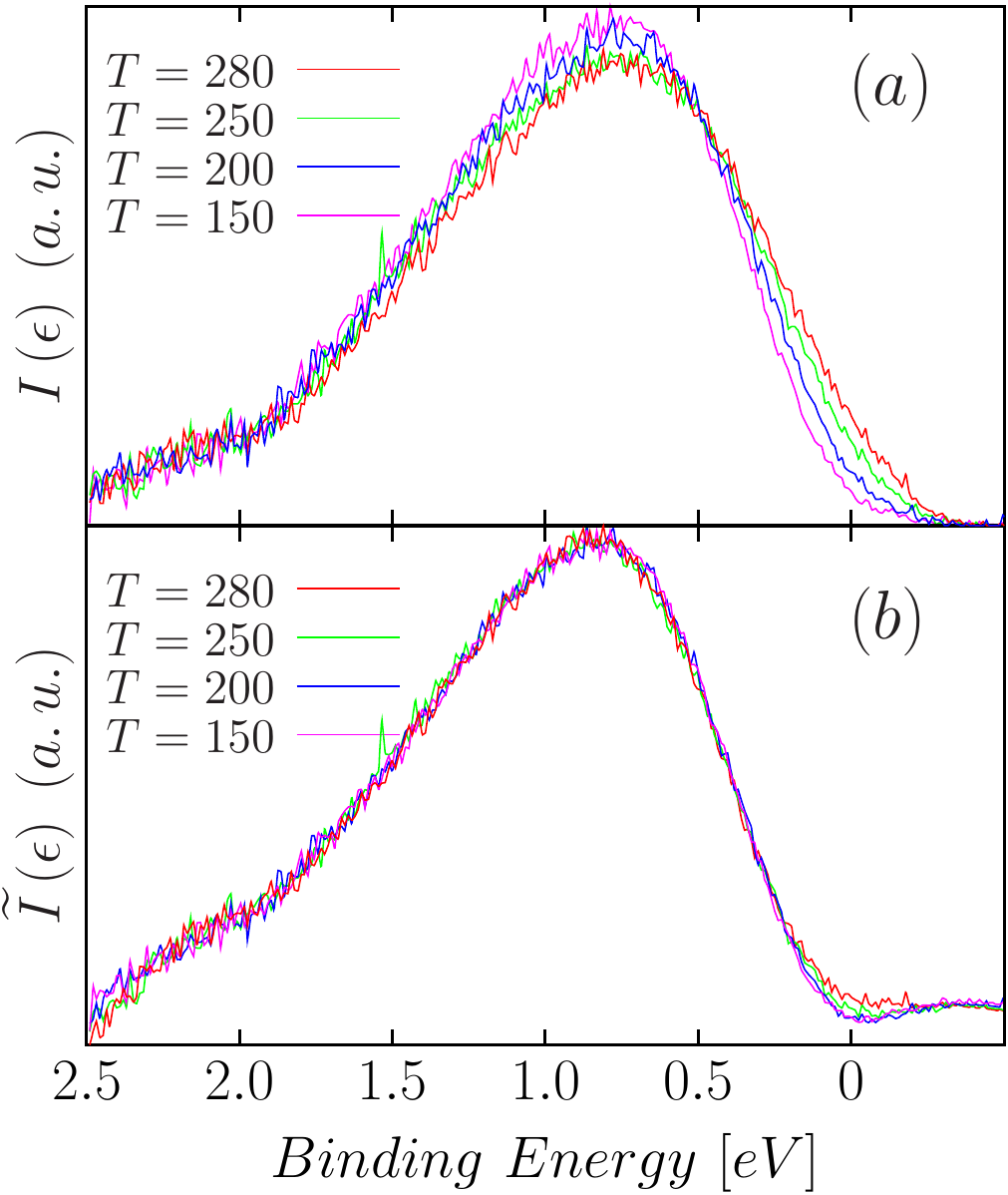}\caption{\label{fig:Fig7}Temperature dependence of (a) photoemission spectra
$I\left(\omega,T\right)$ of a VO$_{2}$ thin film {[}\cite{PhysRevB.69.165104}{]}, and (b) the function
$\widetilde{I}\left(\omega,T\right)$ defined in Eq. \eqref{eq:Delta_phot_em_spec}.
The weak temperature dependence of $\widetilde{I}\left(\omega,T\right)$
explains the sharpness of the isosbestic point and also supports the
importance of electron-phonon interaction.}
\end{figure}

\subsection{\label{sub:Photoemission-spectra-of}Photoemission spectra of VO$_{2}/$TiO$_{2}$
$(001)$ thin films}

Next we apply the method to explain photoemission spectra $I\left(\omega,T\right)$
of VO$_{2}/$TiO$_{2}$ thin films {[}\cite{PhysRevB.69.165104}{]},
which exhibit a marked isosbestic (see Fig.~\ref{fig:Fig7}). The
strong temperature dependence of $I\left(\omega,T\right)$ was seen
{[}\cite{PhysRevB.69.165104}{]} as an indication of strong electron-phonon
coupling. Consequently, $I\left(\omega,T\right)$ was {[}\cite{PhysRevB.69.165104}{]}
convincingly reproduced through the spectral function $A\left(\omega,T\right)$
of the independent boson model through $I\left(\omega,T\right)\propto A\left(\omega,T\right)$.
The temperature dependence of $A\left(\omega,T\right)$ with $A\left(\omega,T\right)=A\left(\omega,0\right)+T\, A_{1}\left(\omega\right)+\mathcal{O}\left(T^{2}\right)$
justifies the following ansatz:
\begin{equation}
I\left(\omega,T\right)=I\left(\omega,0\right)+T\, I_{1}\left(\omega\right)+\mathcal{O}[T^{2}],\label{eq:photo_em}
\end{equation}
 where
\[
I_{1}\left(\omega\right)\simeq\frac{I(\omega,T_{1})-I(\omega,T_{2})}{T_{1}-T_{2}}
\]
is extracted from the data ($T_{1}=150K$ and $T_{2}=200K$). The
ansatz \eqref{eq:photo_em} is confirmed by the relative $T$-independence
(well within measurement accuracy) of the quantity
\begin{eqnarray}
\widetilde{I}\left(\omega,T\right) & = & I\left(\omega,T\right)-T\, I_{1}(\omega)\nonumber \\
 & \simeq & I\left(\omega,T=0\right),\label{eq:Delta_phot_em_spec}
\end{eqnarray}
plotted in Fig.~\ref{fig:Fig7}b. This provides an explanation for
the sharpness of the isosbestic point, and the leading linear temperature
dependence supports the importance of bosonic excitations along the
line of Ref.~{[}\cite{PhysRevB.69.165104}{]}.

\subsection{\label{sub:Phonon-modes-in}Phonon modes in CaCu$_{3}$Ti$_{4}$O$_{12}$}

\begin{figure}[t]
\centering{}\includegraphics[scale=1.1]{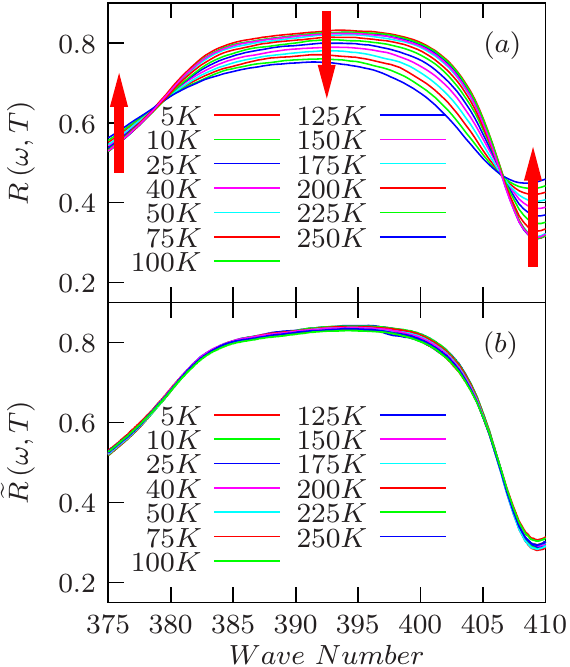}\caption{\label{fig:Fig8}Temperature
  dependence of (a) the reflectivity $R\left(\omega,T\right)$
of CaCu$_{3}$Ti$_{4}$O$_{12}$ {[}\cite{kant2008bdr}{]} and (b) the function $\widetilde{R}\left(\omega,T\right)$.
The weak temperature dependence of $\widetilde{R}\left(\omega,T\right)$
explains the origin of the isosbestic point.}
\end{figure}

We investigate isosbestic points in reflectivity $R\left(\omega,T\right)$
measurements of CaCu$_{3}$Ti$_{4}$O$_{12}$ around phononic excitations
{[}\cite{kant2008bdr}{]}. Again, assuming the leading $T$-dependence
to be linear, we proceed along the lines of the preceding section
and introduce
\[
\widetilde{R}\left(\omega,T\right)=R\left(\omega,T\right)-T\, R_{1}(\omega),
\]
 where
\begin{equation}
R_{1}\left(\omega\right)\simeq\frac{R(\omega,T_{1})-R(\omega,T_{2})}{T_{1}-T_{2}}\label{eq:R1_omega_T}
\end{equation}
 is extracted directly from data. Note that the sensitivity of Eq.~\eqref{eq:R1_omega_T}
on the quality of the available data required the use of a comparably
large temperature difference ($T_{1}=40K$, $T_{2}=175K$). The high
quality of the match in Fig.~\ref{fig:Fig8} explains the sharpness
of the isosbestic point and also shows that a linear temperature dependence
already provides a proper description of the available measurement
data.

\subsection{\label{sec:subPolymer}Isosbestic points in conductive polymers}

Here we analyze the marked isosbestic points in the optical conductivity
of conductive polymers {[}\cite{PhysRevLett.39.1098,C39770000578}{]}
as observed in many spectroelectrochemical experiments for different
doping levels. By  applying an external bias voltage $V$, these experiments
allow for direct control of the doping level of the sample and 
yield the optical conductivity (absorbance) $\sigma(\omega,n)$ resp.
$A(\omega,n)$ as a function of doping $n$. 

\begin{figure}[t]
\centering{}\includegraphics[scale=0.8]{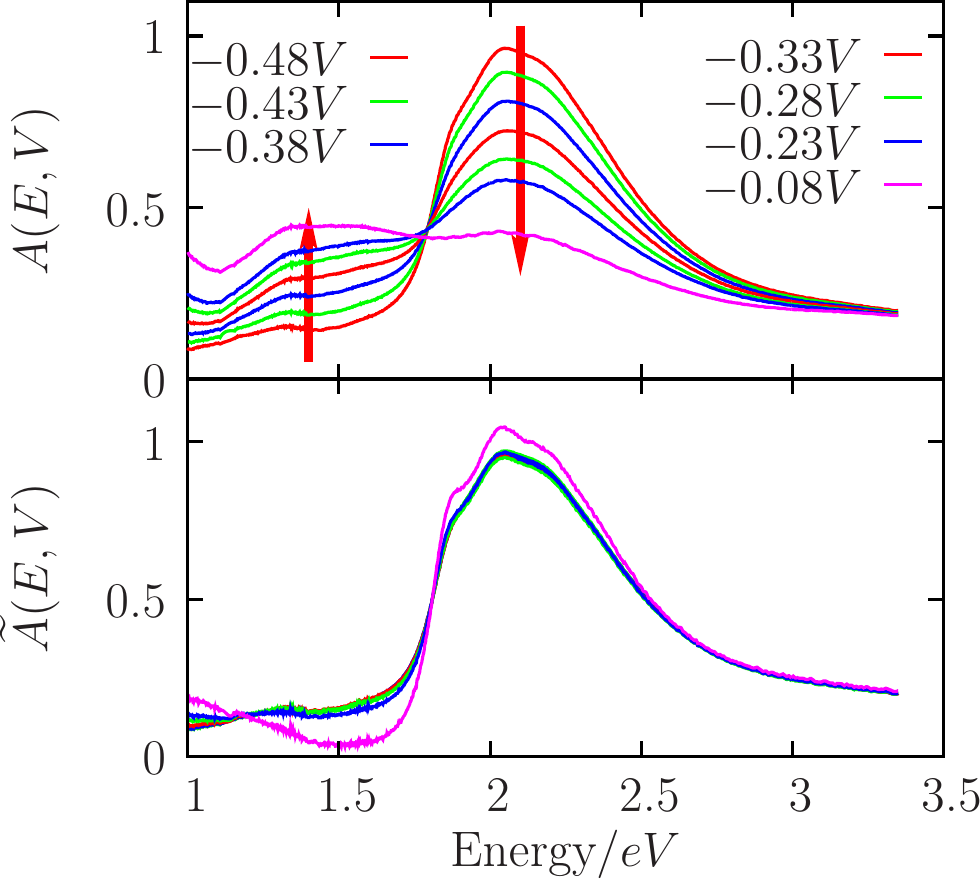}\caption{\label{fig:FigPolymer}The upper plot shows the marked isosbestic
point in spectroelectrochemical experiments for PEDOT. In the lower
plot we show the corresponding curves $\widetilde{A}(\omega,V_{i})$
which (except for $V_{7}=-0.08V$) essentially all collapse onto a
single curve. We included the curves $A(\omega,V_{7})$
and $\widetilde{A}(\omega,V_{7})$ to demonstrate that deviations
become important at more extreme parameter values. In particular, we note
that $A(\omega,V_{7})$ does not contribute to the isosbestic defined
by the other curves. (Data from Ref.~\onlinecite{thomas2002donor}.)}
\end{figure}

To demonstrate the general applicability of our scheme we specifically
consider a subset of the spectroelectrochemical measurements on poly($3$,$4$-ethylenedioxythiophene)
(PEDOT) presented in Ref.~{[}\cite{thomas2002donor}{]}. Note that
we consider here $A(\omega,V)$ instead of $A(\omega,n)$ since $V$
is the original parameter in the measurement. We make the following
linear ansatz for the parameter-dependence of $A(\omega,V)$, expanding
around $V_{0}=-0.48V$ 
\begin{equation}
A(\omega,V)=A(\omega,V_{0})+(V-V_{0})A_{1}(\omega),\label{eq:SpecElchemAnsatz}
\end{equation}
where 
\[
A_{1}(\omega)\simeq\frac{A(\omega,V_{1})-A(\omega,V_{2})}{V_{1}-V_{2}}
\]
with $V_{1}=-0.48$ and $V_{2}=-0.38$. Again we plot both $A(\omega,V)$
and $\widetilde{A}(\omega,V)\equiv A(\omega,V)-(V-V_{0})A_{1}(\omega)$
in Fig.~\eqref{fig:FigPolymer}, verifying the applicability of our
general scheme and in particular of Eq. \eqref{eq:SpecElchemAnsatz}. 
Our analysis thus reveals that the seemingly complicated behavior
of $A(E,V)$ in Fig.~\ref{fig:FigPolymer} is in fact due to a linear
voltage dependence.

\subsection{\label{sub:LAMnO3}Isosbestic point in the dielectric function of LaMnO$_{3}$}

\begin{figure}[t]
\centering
\includegraphics[scale=0.8]{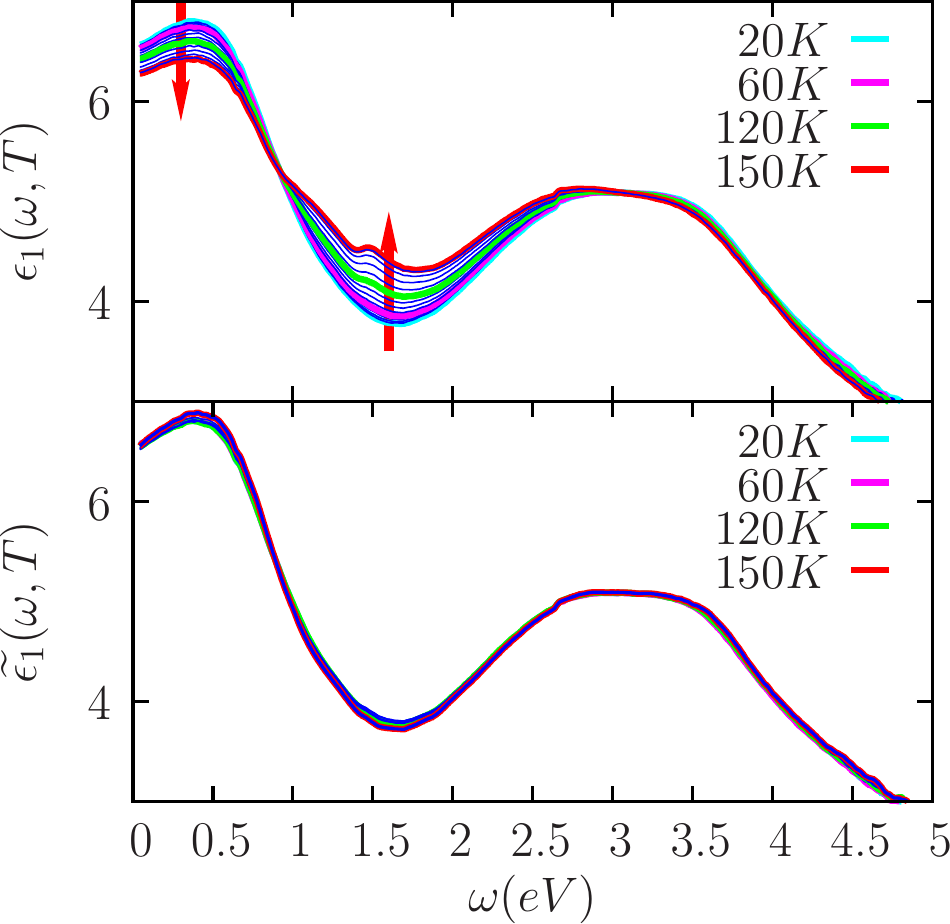}

\caption{\label{fig:Kovaleva}Temperature variation of the real part of the
dielectric function spectra $\epsilon_{1}(\omega,T)$ of LaMnO$_{3}$
between $20$K and $140$K in $10$K increments.  (Data from Ref.~\onlinecite{PhysRevB.81.235130}.)}
\end{figure}

Here we address the conspicuous isosbestic points in the dielectric function
$\epsilon_{1}(\omega,T)$ of LaMnO$_{3}$ for different temperatures
{[}\cite{PhysRevB.81.235130}{]} (see Fig.~\ref{fig:Kovaleva}). We apply the same analysis as before, i.e., 
we use the following Sommerfeld-type
ansatz for the temperature-dependence of $\epsilon_{1}(\omega,T)$
\begin{equation}
\epsilon_{1}(\omega,T)=\epsilon_{1}(\omega,0)+T^{2}\epsilon_{1}^{(1)}(\omega)+\mathcal{O}[T^{3}],\label{eq:Kovaleva}
\end{equation}
where ($T_{1}=20\text{K}$, $T_{2}=90\text{K}$) 
\begin{equation}
\epsilon_{1}^{(1)}(\omega)\simeq\frac{\epsilon_{1}(\omega,T_{1})-\epsilon_{1}(\omega,T_{2})}{T_{1}^{2}-T_{2}^{2}}.\label{eq:Kovaleva_deriv}
\end{equation}
as in Eq.~\eqref{eq:approx_Q1}. Again, the validity of Eq.~\eqref{eq:raman_ch_exp}
is tested via the $T$-independence of the quantity
\[
\widetilde{\epsilon_{1}}(\omega,T)=\epsilon_{1}(\omega,T)-T^{2}\epsilon_{1}^{(1)}(\omega).
\]
This reveals (i) that the origin of the optical response is dominated by electronic
excitations that are quadratic in $T$, and (ii) that the temperatures are still moderately 
small as compared to the electronic low-energy scales of the system.

\section{\label{sec:FK}Isosbestic points in the optical conductivity of the
Falicov-Kimball model}

\subsection{Model and optical conductivity in DMFT}

The Falicov-Kimball model {[}\cite{falicov1969simple,RevModPhys.75.1333}{]}
describes itinerant electrons ($c$-electrons) interacting with localized
electrons ($f$-electrons) by a local Coulomb repulsion. The Hamiltonian
is given by
\begin{eqnarray}
H & = & -\sum_{i,j}t_{i,j}c_{i}^{\dagger}c_{j}-\mu\sum_{i}\left(c_{i}^{\dagger}c_{i}+f_{i}^{\dagger}f_{i}\right)\label{eq:hamil}\\
 &  & +E_{f}\sum_{i}f_{i}^{\dagger}f_{i}+U\sum_{i}\left[f_{i}^{\dagger}f_{i}-\frac{1}{2}\right]\left[c_{i}^{\dagger}c_{i}-\frac{1}{2}\right].\nonumber
\end{eqnarray}
It can be viewed as a Hubbard model with spin-dependent hopping and
two different chemical potentials. The symbols $c_{i}^{\dagger}$
($f_{i}^{\dagger}$) and $c_{i}$ ($f_{i}$) denote the itinerant
(localized) electron creation and annihilation operators. The chemical
potential $\mu$ constrains the total number of $c$-electrons, while
$E_{f}$ is an orbital energy of the localized electrons. Here $U$
is an on-site repulsion between the two electron species.

\begin{figure}[t]
\centering{}\centering{}\includegraphics[width=0.82\columnwidth]{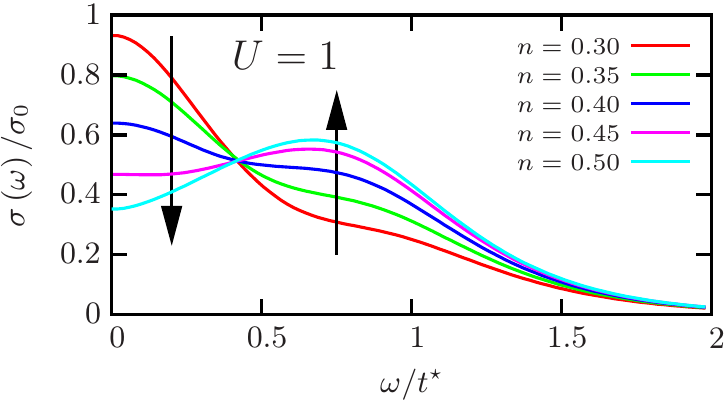}\caption{\label{fig:Fig1}Isosbestic point in the optical conductivity of the
Falicov-Kimball model at $T=0$ {[}\cite{RevModPhys.75.1333}{]} (Gaussian
DOS {[}$U_{c}\simeq1.5${]} with density $n$ of mobile and density
$w_{1}=1/2$ of immobile particles). Spectral weight is transferred
from low frequencies to the Hubbard peak as electron-density is increased,
producing an isosbestic point at medium frequencies.}
\end{figure}

At low temperatures, the optical conductivity of the Falicov-Kimball
model with the Gaussian density of states (DOS) is known to exhibit
a distinct isosbestic point {[}\cite{RevModPhys.75.1333}{]} when plotted
for various $c$-electron densities (Fig.~\ref{fig:Fig1}). \sout{}
In the following we will discuss the existence and sharpness of isosbestic
points and concentrate on the ungapped system at zero temperature
and use the semi-elliptic DOS $\rho(\epsilon)=(2\pi t)^{-1}\sqrt{4t^{2}-\epsilon^{2}}$
with $t=1$.

Within dynamical mean-field theory (DMFT), the model can be solved
exactly for the Green function $G(\omega)$ or the self-energy $\Sigma(\omega)$
owing to the simplicity of the impurity problem {[}\cite{brandt1989tac}{]}:
\begin{eqnarray}
G(\omega) & = & \frac{1-w_{1}}{\mathcal{G}_{0,c}^{-1}(\omega)}+\frac{w_{1}}{\mathcal{G}_{0,c}^{-1}(\omega)-U},\label{eq:imp_prob}
\end{eqnarray}
 where $\mathcal{G}_{0,c}^{-1}(\omega)=\Sigma(\omega)+G^{-1}(\omega)$
denotes the Green function of the effective medium (Weiss field).
The weight $w_{1}$ equals the average $f$-electron concentration
which will be set to $1/2$ in the following. Due to particle-hole
symmetry, it is sufficient to restrict the treatment to $c$-electron
densities $n\leq1/2$, i.e., $\mu\leq0$.

The optical conductivity in DMFT is essentially given by the particle-hole
bubble {[}\cite{jarrell1995oci}{]}, which on the real axis corresponds
to
\begin{eqnarray}
\sigma(\omega) & = & \sigma_{0}\int_{-\infty}^{\infty}d\epsilon\rho(\epsilon)v(\epsilon)^{2}\int d\omega^{\prime}A_{\epsilon}\left(\omega^{\prime}\right)A_{\epsilon}\left(\omega^{\prime}+\omega\right)\nonumber \\
 &  & \times\frac{f\left(\omega^{\prime}\right)-f\left(\omega^{\prime}+\omega\right)}{\omega},\label{eq:sigma_omega}\\
A_{\epsilon}(\omega) & = & -1/\pi\, \text{Im}\left[\omega+\mu-\Sigma(\omega)-\epsilon\right]^{-1},\label{eq:spec_dens}
\end{eqnarray}
 where $A_{\epsilon}(\omega)$ denotes the spectral density and $\rho(\epsilon)$
the density of states (DOS). We fix the velocity function $v(\epsilon)$
according to the treatment in Ref.~{[}\cite{blumer2003tpc}{]}.

\begin{figure}
\begin{centering}
\includegraphics[width=0.82\columnwidth]{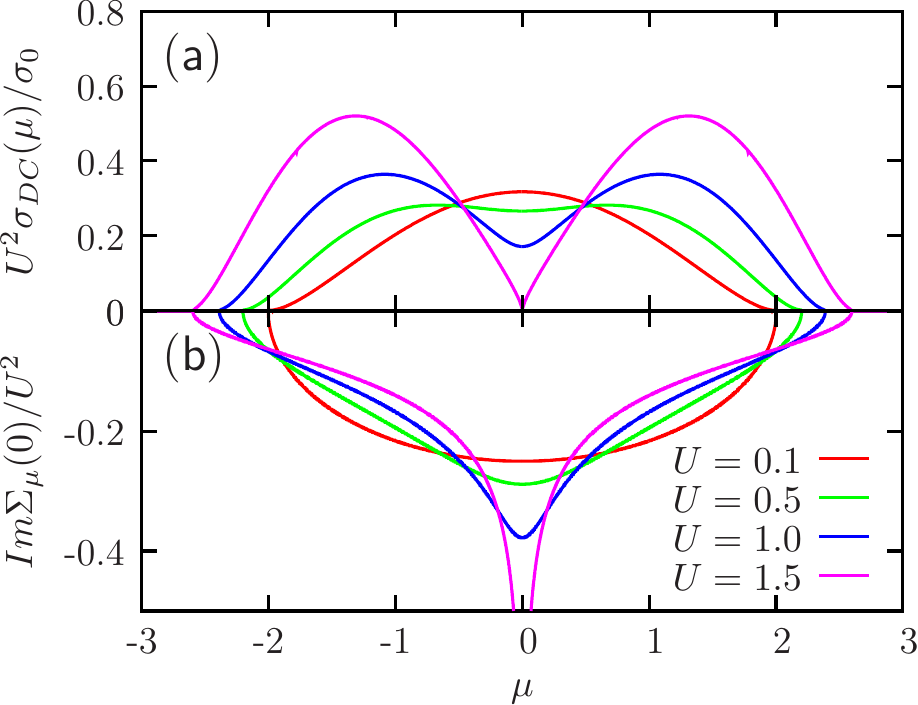}
\par\end{centering}

\caption{\label{fig:Fig2}Doping dependence of (a) the rescaled DC conductivities
$U^{2}\sigma_{DC}\left(\mu\right)$ and (b) self-energies $\text{Im}\Sigma_{\mu}(0)/U^{2}$
for the semi-elliptic DOS. With growing interaction $\text{Im}\Sigma_{\mu}(0)$
becomes increasingly peaked with an extremum at half-filling ($\mu=0$).
Therefore the scattering rate  $\tau^{-1}\propto-\text{Im}\Sigma(0)$ becomes
maximal at half-filling. The metal-insulator transition at $U=2$
is distinguished by a vanishing DC conductivity and a diverging scattering
rate $\tau^{-1}$. For sufficiently high, yet finite $U$ this generic
property of the FK model leads to the formation of minima in $\sigma_{DC}\left(\mu\right)$
around for $\mu=0$. This corresponds to $\partial\sigma_{DC}(\mu)/\partial\mu<0$.
Note that there exists a one-to-one mapping between $\mu$ and the
carrier density $n$. In Appendix \ref{sec:Shape-invariance} we show
that $\Sigma_{\mu}(\omega)=\Sigma(\omega+\mu)$.}
\end{figure}

\begin{figure}[t]
\centering{}\includegraphics[width=0.82\columnwidth]{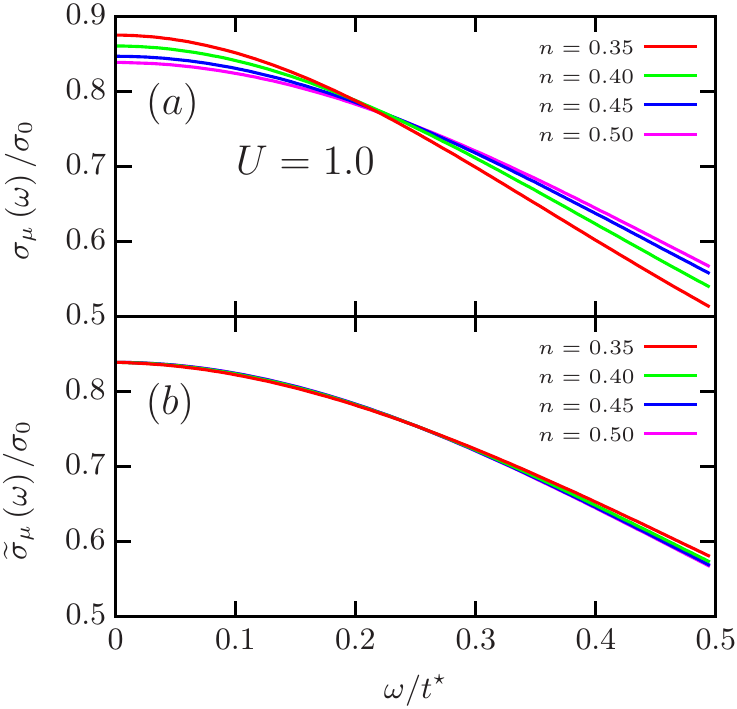}\caption{\label{fig:Fig3}Optical conductivity (a) $\sigma(\omega,\mu\left(n\right))$
and (b) the quantity $\widetilde{\sigma}(\omega,\mu\left(n\right))$
for the Falicov-Kimball model with the semi-elliptic DOS. The curves
$\widetilde{\sigma}(\omega,\mu\left(n\right))$ in (b) essentially
lie on top of $\sigma\left(\omega,\mu=0\right)$ for the substantial
density interval $n\in\left[0.35\ldots0.5\right]$. Thus, the parameter
dependence in $\sigma(\omega,\mu\left(n\right))$ essentially stems
from the first contribution, which is accompanied by an isosbestic
point. The second derivative of $\sigma\left(\omega,\mu\right)$ w.r.t.
$\mu$, which is needed for the calculation of $\widetilde{\sigma}\left(\omega,\mu\right)$,
is obtained from the exact solution {[}\cite{PhysRevB.45.2267}{]}.}
\end{figure}

\subsection{\label{sec:Exist-1}Existence of isosbestic points in $\sigma(\omega)$}

The formation of isosbestic points in the optical conductivity $\sigma(\omega)$
when plotted for different doping levels of the c-electrons in Fig.~\ref{fig:Fig1}
can easily be deduced on physical grounds, viewing the Falicov-Kimball
model as a disordered medium in which spinless electrons interact
with localized impurities. In this picture the scattering rate $\tau^{-1}(\mu)\propto-\text{Im}\Sigma_{\mu}(\omega=0)$
is monotonically related to the c-electron density through the chemical
potential $\mu$, and reaches a maximum at half-filling (see Fig.~\ref{fig:Fig2}b).
Since $\tau^{-1}(0)$ diverges at the metal-to-insulator transition,
it is always possible to suppress the DC conductivity of the half-filled
FK model simply by increasing $U$. Thus, in the intermediate $U$-regime
$\sigma_{DC}(U,n)$ will have a minimum or a dip {[}\cite{RevModPhys.75.1333}{]}
around half-filling which means $\partial\sigma_{DC}/\partial n<0$
for $n<1/2$ (see Fig.~\ref{fig:Fig2}a). Besides the mathematical
reasoning in Appendix~\ref{app:AppA}, a positive sign of $\partial\sigma(\omega,n)/\partial n$
for non-zero frequencies and for sufficiently strong $U$ can be established
by taking into account the formation of the well-known charge-transfer
or Hubbard peak at $\omega_{C}(U,n)\simeq U$ (see Fig.~\ref{fig:Fig1}),
which is a property of the optical conductivity of correlated systems.
Representing a robust indicator of the increasing correlation between
$f$- and $c$-electrons, the peak becomes increasingly pronounced
as the system is doped towards half-filling. For sufficiently strong
$U$ we thus have $\partial\sigma\left(\omega_{C}\right)/\partial n>0$
and $\partial\sigma_{DC}/\partial n<0$, and as a consequence there
exists a crossing point $\omega^{\star}(n,U)$ (c.f. Eq.~\eqref{crossing})
with $0<\omega^{\star}<\omega_{C}$ and $\partial\sigma\left(\omega^{\star},n\right)/\partial n=0$
as depicted in Fig.~\ref{fig:Fig1}. The formation of isosbestic
points is thus an effect of the correlation associated transfer of
spectral weight from low to higher frequencies as the system is doped
towards half-filling. The argument does not depend on the specifics
of the density of states under consideration and thus applies to Fig.~\ref{fig:Fig1}
(Gaussian DOS) as well as to Fig.~\ref{fig:Fig3} (semi-elliptic
DOS).

\subsection{\label{sec:Sharpness}Sharpness of isosbestic points in $\sigma(\omega)$}

\begin{figure}[t]
\centering{}\centering{}\includegraphics[width=0.82\columnwidth]{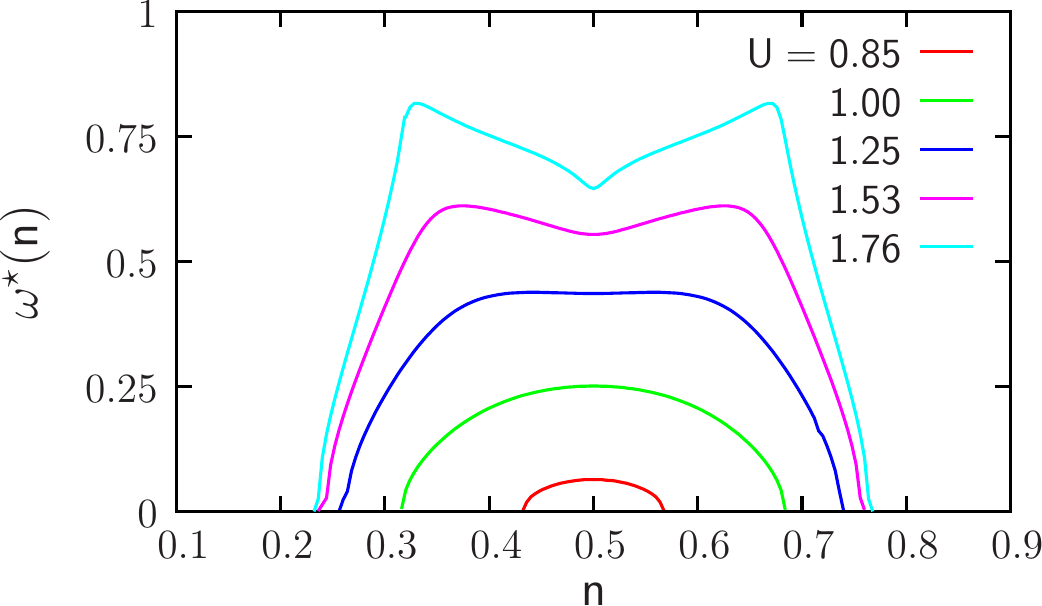}\caption{\label{fig:Fig4}Isosbestic points $\omega^{\star}\left(n,U\right)$
of the ungapped system using the semi-elliptic density of states.
The symmetry around half-filling is a consequence of particle-hole
symmetry.}
\end{figure}

The symmetric density dependence of the crossing frequency $\omega^{\star}(n)$
depicted in Fig.~\ref{fig:Fig4} reflects the important role of particle-hole
symmetry for locally exact isosbestic points in $\sigma\left(\omega,n\right)$.
Because of the existence of the one-to-one correspondence between
the chemical potential $\mu$ and the carrier density $n$ in the
ungapped system both parameters are equivalent. The chemical potential,
however, represents the more convenient parameter because of the simple
$\mu$-dependence of the self-energy $\Sigma(\omega)$ (see Appendix
\ref{sec:Shape-invariance}) and of the spectral densities $A_{\epsilon}(\omega)$
in Eq.~\eqref{eq:sigma_omega}. Thus, concentrating on ungapped systems,
we expand around half-filling, i.e. $\mu=0$, taking $\mu$ as a small
parameter
\begin{equation}
\sigma\left(\omega,\mu\right)=\sigma\left(\omega,0\right)+\frac{1}{2}\mu^{2}\frac{\partial^{2}}{\partial\mu^{2}}\sigma\left(\omega,\mu=0\right)+\mathcal{O}\left(\mu^{4}\right).\label{eq:sigma_expnd}
\end{equation}
With Eq.~\eqref{eq:gen_expand_interpret} we are directly led to
the conclusion that the first two terms constitute an exact isosbestic
point, while corrections of order $\mathcal{O}\left(\mu^{4}\right)$
introduce the $\mu$ dependence into $\omega^{\star}\left(\mu\right)$
(or $\omega^{\star}(n)$). According to Eq.~\eqref{eq:Delta_f} we
therefore introduce
\begin{eqnarray*}
\widetilde{\sigma}\left(\omega,\mu\right) & = & \sigma\left(\omega,\mu\right)-\frac{1}{2}\mu^{2}\frac{\partial^{2}}{\partial\mu^{2}}\sigma\left(\omega,\mu=0\right)\\
 & = & \sigma\left(\omega,\mu=0\right)+\mathcal{O}[\mu^{4}],
\end{eqnarray*}
and test its approximate $\mu$ independence in a vicinity of $\omega^{\star}\left(\mu=0\right)$.
This is confirmed graphically in Fig.~\ref{fig:Fig3}; the obvious
agreement with the reference curve can be seen as a measure of the
weak influence of higher order contributions. This provides an especially
simple and meaningful interpretation of the approximate isosbestic
points in $\sigma\left(\omega,n\right)$: Namely, plots of the conductivity
showing this behavior exhibit a density dependence which can be treated
in perturbation theory for the parameters under consideration.

\section{\label{sec:Hubbard}Isosbestic points in the spectral function of
the Hubbard model}

We now apply our approach to the prominent isosbestic point in the
spectral function $A\left(\omega,U\right)$ of the particle-hole symmetric
Hubbard model within DMFT for which, employing symmetry arguments,
the crossing frequency was shown {[}\cite{eckstein2007ips}{]} to
obey the following dependence on $U$,
\begin{equation}
\omega^{\star}\left(U\right)=c_{0}+c_{1}U^{2}+\mathcal{O}[U^{3}].\label{eq:hubb_omega_star}
\end{equation}
For a symmetric DOS with van-Hove singularities at the band-edges
(at $\pm b$) like the semi-elliptic DOS, Eq.~\eqref{eq:hubb_omega_star}
takes the more specific form $\omega^{\star}\left(U\right)=b+c_{1}U^{2}+\mathcal{O}(U^{3})$.

\begin{figure}
\centering{}\centering{}\includegraphics[width=0.82\columnwidth]{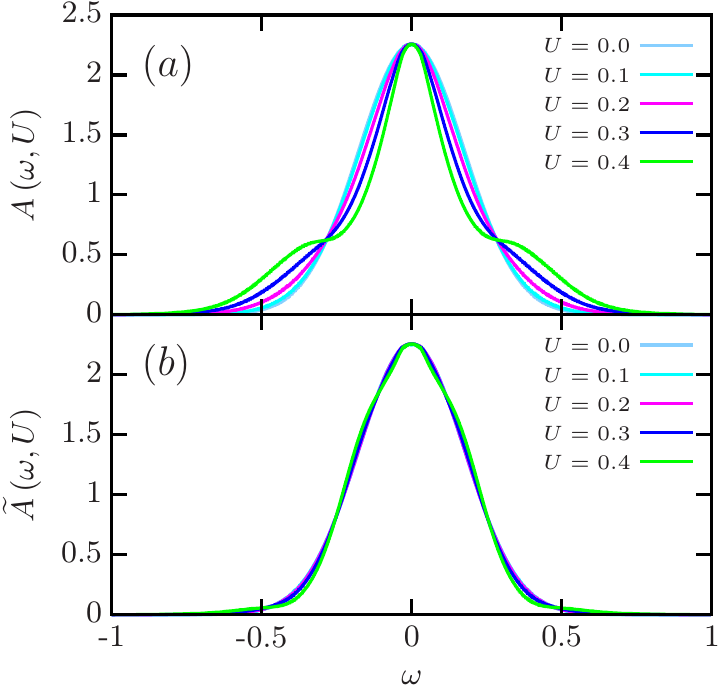}\caption{\label{fig:Fig5}Numerical renormalization group  calculation {[}\cite{RevModPhys.80.395,nrgcode2}{]}
of the spectral function for the half-filled Hubbard model (a) using
the Gaussian DOS $\rho(\epsilon)=4/\sqrt{\pi}\exp[-(4\epsilon)^{2}]$.
The quantity $\widetilde{A}\left(\omega,U\right)$ shows practically
no $U$-dependency up to $U_{1}\simeq0.35$ (b). The deviations from
the locally exact isosbestic point at higher $U$ stem from higher
order perturbations and become noticeable only above $U_{1}$.}
\end{figure}

From the infinite slope of the spectral function $A\left(\omega,0\right)$
at $\omega=\omega^{\star}\left(0\right)$ it then follows that, in
the limit $U\rightarrow0$, one can never obtain a sharp crossing
\textit{point} defined by the coordinates $\left(x(U),y(U)\right)$
with $x(U)=\omega^{\star}\left(U\right)$ and $y(U)=A(\omega^{\star}(U),U)$.
In order to avoid this additional mathematical complication, let us
here consider the Gaussian DOS with infinite bandwidth, since this
DOS is a smooth function so that a weak parameter dependence of $\omega^{\star}(U)$
always translates to a sharp isosbestic point. From particle-hole
symmetry, we have the following perturbative expansion
\begin{equation}
A\left(\omega,U\right)=\sum_{n\geq0}U^{2n}A_{2n}(\omega)\label{eq:SpecDensExpand}
\end{equation}
 with
\begin{equation}
A_{2}(\omega)=-\frac{1}{\pi}\text{Im}\left\{ \int d\epsilon\,\rho(\epsilon)\frac{\Sigma_{2}(\omega)}{\left[\omega+i0^{+}-\epsilon\right]^{2}}\right\} .\label{eq:spec_fun_A2}
\end{equation}
 Here $\Sigma_{2}(\omega)$ denotes the self-energy in second order
perturbation theory {[}\cite{schweitzer1991wct}{]}  in $U$. We thus
obtain from Eqs.~(\ref{eq:Delta_f}-\ref{eq:Delta_f3})
\begin{eqnarray}
\widetilde{A}\left(\omega,U\right) & = & A\left(\omega,U\right)-U^{2}A_{2}(\omega)\label{eq:Delta_spec_fun}
\end{eqnarray}
 and again obtain $\widetilde{A}\left(\omega,U\right)\simeq A\left(\omega,0\right)$,
as can be seen from Fig.~\ref{fig:Fig5}.

\section{\label{sec:Conclusion}Conclusion}

In this paper we developed a framework which allows one to extract
information about a correlated system by analyzing the sharpness of
an approximate isosbestic point in a physical quantity $f(x,p)$.
Our central result is given by Eq.~\eqref{eq:gen_expand_interpret}
which corresponds to a straightforward generalization of Eq.~\eqref{absorbance},
describing exact isosbestic points {[}\cite{Scheibe1937,cohen1962ip}{]}.
We established a direct connection between the sharpness of isosbestic
points and the existence of a small parameter $\delta_{p}=p-p_{0}$,
which can be used to extract the leading parameter dependence of
the system in the parameter interval under consideration.

For example, our analysis of the Raman response of
HgBa$_{2}$CuO$_{4+\delta}$ showed that it has a quadratic temperature
dependence only for optimal doping, which raises the question of
whether other dynamical quantities exhibit such behavior as well.
Another interesting question is whether other (optimally doped)
cuprate compounds have similar isosbestic points in their Raman
spectra. Furthermore, we deduced a nonobvious linear voltage
dependence for the absorbance of the conductive polymer PEDOT in a
wide frequency range.

In summary, if a sharp isosbestic point is found in experimental
data, an analysis as in Sec.~III~A-E should be performed to extract
the leading parameter dependence of the family of curves.

\section{\label{sec:Acknowledgments}Acknowledgments}

We are thankful to J. Deisenhofer, I.~{K\'{e}zsm\'{a}rki}, A. Loidl,
and Y. Tokura for useful communication. This work was supported by
the Deutsche Forschungsgemeinschaft through TRR~80.

\appendix
\appendix

\section{\label{sec:Size-of-the}Size of the crossing region around locally
exact isosbestic points}

We quantify the influence of the higher-order contributions, i.e.,
the terms responsible for the deviations from $\XSTAR(p)=const$.
For this we investigate the $p$-dependence of an isosbestic point
at $(\XSTAR(p),p)$, with $\XSTAR(p)$ given implicitly by Eq.~\eqref{crossing}.
We will estimate the size of the region where the curves of $f(x,p)$
cross, e.g., for \begin{subequations}
\begin{align}
x & =\XSTAR(p_{0})+\delta x\,,\\
p & =p_{0}+\delta p\,,
\end{align}
 \end{subequations}in the vicinity of a certain parameter value $p_{0}$.

We first approximate $f(x,p)$ near $p_{0}$ by a function that is
linear in $p$,
\begin{align}
f_{\text{approx}}(x,p) & =f(x,p_{0})+\delta p\, F_{1}(x,p_{0})\,,\label{eq:gen_expand_interpret_tmpK}
\end{align}
 where $F_{n}(x,p)=d^{n}f(x,p)/dp^{n}$. Equation \eqref{eq:gen_expand_interpret_tmpK}
is exact at $p_{0}$: $f_{\text{approx}}(x,p_{0})$ $=$ $f(x,p_{0})$.
As in the two-fluid model that was discussed in the introduction,
$f_{\text{approx}}(x,p)$ has an exact isosbestic point at $(\XSTAR(p_{0}),p_{0})$
because $F_{1}(x,p_{0})$ vanishes there.

Next we determine the difference between the function $f(x,p)$ and
the linear approximation $f_{\text{approx}}(x,p)$ is. For this purpose
we first expand $F_{1}(x,p)$ near $\XSTAR(p)$ for some fixed $p$
and obtain
\begin{align}
F_{1}(x,p) & =F_{1}(\XSTAR(p),p)+\sum_{n=1}^{\infty}\frac{[x-\XSTAR(p)]^{n}}{n!}\left.\frac{\partial^{n}F_{1}}{\partial x^{n}}\right|_{x=\XSTAR(p)}.\label{eq:higher_order_expK}
\end{align}
By virtue of the definition of $\XSTAR(p)$ {[}Eq.~\eqref{crossing}{]},
the $x$-independent term vanishes, i.e., $F_{1}(\XSTAR(p),p)$
$=$ $0$, leading to
\begin{align}
F_{1}(x,p) & =c(p)\,[x-\XSTAR(p)]+O([x-\XSTAR(p)]^{2})\,,\label{eq:linearF1K}
\end{align}
 with $c(p)$ $=$ $\partial F_{1}(x,p)/\partial x|_{x=\XSTAR(p)}$.

The height of the crossing region is thus given by
\begin{align}
\delta f(x,p) & =f(x,p)-f_{\text{approx}}(x,p)\nonumber \\
 & =\frac{\delta p^{2}}{2}\, F_{2}(x,p_{0})+\frac{\delta p^{3}}{6}\, F_{3}(x,p_{0})+O(p^{4})\,.\label{eq:Delta_fK}
\end{align}
 Using Eq.~\eqref{eq:linearF1K} we can now express the higher derivatives
as ($n\geq0$)
\begin{multline}
F_{n+1}(x,p)=c^{(n)}(p)\,[x-\XSTAR(p)]\\
-\sum_{k=1}^{n}\binom{n}{k}c^{(n-k)}(p)\,\XSTAR^{(k)}(p)\,,
\end{multline}
where $c^{(n)}(p)=\partial^{n}F_{1}(x,p)/\partial^{n}x|_{x=\XSTAR(p)}$.

So far the parameter value $p$ has not been specified. We now suppose
that $\XSTAR(p)$ has an extremum at $p_{0}$, in which case we term
$(\XSTAR(p_{0}),p_{0})$ a first-order isosbestic point. From $\XSTAR^{\prime}(p_{0})$
$=$ $0$ it then follows that
\begin{align}
\delta f & =\frac{c^{\prime}(p)}{2}\,\delta p^{2}\,\delta x+O(\delta p^{3}\delta x^{0})\,.
\end{align}
 If instead $\XSTAR(p)$ has a saddle point at $p_{0}$ (second-order
isosbestic point), then furthermore $\XSTAR''(p_{0})$ $=$ $0$,
yielding
\begin{align}
\delta f & =\left(\frac{c^{\prime}(p)}{2}\,\delta p^{2}+\frac{c^{\prime\prime}(p)}{6}\,\delta p^{3}\right)\,\delta x+O(\delta p^{4}\delta x^{0})\,,
\end{align}
 and similar for higher-order isosbestic points. We thus conclude
that an $n$th-order isosbestic point, for which the first $n$ derivatives
of $\XSTAR(p)$ vanish at $p_{0}$, will have a narrow crossing region
\begin{align}
\delta f & =\delta x\,\sum_{k=2}^{n+1}\frac{1}{k!}\, c^{(k-1)}(p)\,\delta p^{k}+O(\delta p^{n+2}\delta x^{0})\,,\\
 & =O(\delta p^{n+2})\text{~~if~~}\delta x=0
\end{align}
 i.e., the first $n+1$ powers of $\delta p$ are suppressed by a
factor $\delta x$, and the contribution of order $\delta x^{0}$
is suppressed by a factor $\delta p^{n+2}$. The last equation applies
if we consider the height of the crossing region $\delta f(\XSTAR(p_{0}),p)$
for various parameters $p$ near $p_{0}$, and this height is only
of order $O(\delta p^{n+2})$ for an $n$th order crossing point.
We conclude that the isosbestic point $\XSTAR(p)$ of a function $f(x,p)$
will be particularly sharp in the vicinity of a extremum or higher-order
stationary point of $\XSTAR(p)$.

\section{\label{sec:SharpnessS} Quantification of the sharpness of isosbestic points}

It is not possible to introduce an unambiguous measure for the sharpness
of an approximate isosbestic point because of its explicit dependence
on the magnification of the crossing region. This means that the observed
sharpness always depends on the specifics of plotting. For fixed plot
parameters, i.e., for $f(x,p)$ plotted for $x\in[x_{a},x_{b}]$ and
$p\in[p_{1},p_{n}]$, one can, however, define a measure for the relative
sharpness of a crossing region through the ratio between the area
of the plot and the area of the crossing region. For this we define
the vertical width of the plot as
\begin{equation}
\Delta f=\max_{i,x_{1},x_{2}}\left|f(x_{1},p_{i})-f(x_{2},p_{i})\right|\label{eq:Deltaf}
\end{equation}
with $i=1,\ldots,n$ and $x_{1,2}\in[x_{a},x_{b}]$ and correspondingly
the vertical width of the crossing region as
\begin{equation}
\Delta f^{\star}=\max_{i,j}\left|f_{i}^{\star}-f_{j}^{\text{\ensuremath{\star}}}\right|,\label{eq:DeltaFstar}
\end{equation}
where $f_{i}^{\star}=f(x^{\star}(p_{\text{i}}),p_{i})$. With the
horizontal width of the crossing region defined by
\begin{equation}
\Delta x^{\star}=\max_{i,j}\left|x^{\star}(p_{i})-x^{\star}(p_{j})\right|\label{eq:Deltaxstar}
\end{equation}
one can introduce a quantity $S$ through
\begin{equation}
S=\left[\frac{\Delta f^{\star}}{\Delta f}\frac{\Delta x^{\star}}{\Delta x}\right]^{-1}\label{eq:Sharpn}
\end{equation}
where $\Delta x=|x_{a}-x_{b}|$ which quantifies the sharpness, i.e.,  $S=\infty$ for an exact
isosbestic points and $S\gg1$ for sufficiently sharp isosbestic points.
We stress the explicit dependence of $S$ on $\Delta x$ and the range of parameters $[p_{1},p_{n}]$
unless $S=\infty$. For finite $S$ the qualification {}``sharp''
should thus be considered a relative term and be used with caution.

\section{\label{sec:Exist}Existence of isosbestic points in $\sigma(\omega)$}

First we investigate the system specific mechanisms that lead to isosbestic
points in $\sigma\left(\omega,n\right)$. To allow for a rigorous
analysis, we restrict ourselves only to cases in which there exists
at most one isosbestic point. This applies to the results {[}\cite{RevModPhys.75.1333}{]}
obtained for the Gaussian DOS as well as to our calculations for the
semi-elliptic DOS. Due to particle-hole symmetry one has $\sigma\left(\omega,\mu\right)=\sigma\left(\omega,-\mu\right)$.\sout{
} According to Eq.~\eqref{crossing}, isosbestic points are determined
by
\begin{equation}
\frac{\partial\sigma(\omega^{\star}(n),\, n)}{\partial n}\bigg|_{\omega^{\star}(n)}=0.\label{eq:cond_n}
\end{equation}
In view of the one-to-one mapping between the chemical potential $\mu$
and the carrier density $n$ in the ungapped system, which is a direct
consequence of the Green function $G(\omega,\mu)$ being a function
only of $\omega+\mu$ (see Appendix \ref{sec:Shape-invariance}),
this is equivalent to
\begin{equation}
\frac{\partial\sigma(\omega,\,\mu)}{\partial\mu}\bigg|_{\omega^{\star}(\mu)}=0.\label{eq:cond_mu}
\end{equation}
Since $\ensuremath{\Sigma\left(\omega\right)}$ is also only a function
of $\omega+\mu$ (see Appendix \ref{sec:Shape-invariance}), we obtain
\begin{align}
\frac{\partial\sigma(\omega,\,\mu)}{\partial\mu} & =\sigma_{0}\int d\epsilon\rho(\epsilon)v(\epsilon)^{2}\int d\omega^{\prime}\frac{\partial}{\partial\omega^{\prime}}\left[A_{\epsilon}\left(\omega^{\prime}\right)A_{\epsilon}\left(\omega^{\prime}+\omega\right)\right]\nonumber \\
 & \times\frac{f\left(\omega^{\prime}\right)-f\left(\omega^{\prime}+\omega\right)}{\omega}.\label{eq:derivocd1}
\end{align}
Integration by parts and using the property $\partial f(\omega)/\partial\omega=-\delta(\omega)$
of the Fermi function gives
\begin{equation}
\frac{\partial}{\partial\mu}\sigma\left(\omega,\mu\right)=\frac{\sigma_{0}}{\omega}\int_{-\infty}^{\infty}d\epsilon\widetilde{\rho}\left(\epsilon\right)A_{\epsilon}\left(0\right)\left[A_{\epsilon}(\omega)-A_{\epsilon}\left(-\omega\right)\right].\label{eq:cond_deriv}
\end{equation}
at $T=0$.

In Appendix~\ref{app:AppA} we show that for less than half-filling,
the derivative $\partial\sigma(\omega,\mu)/\partial\mu$ is positive
around a $\omega_{0}>0$ and never becomes negative in the high frequency
limit. Since we assume the existence of at most one isosbestic point,
it follows that it is exclusively determined by the $\mu$-dependence
of the DC conductivity $\sigma_{DC}\left(\mu\right)=\sigma\left(\omega=0,\mu\right)$:
\begin{equation}
\frac{\partial}{\partial\mu}\sigma_{DC}(\mu)<0\label{eq:exist_condition}
\end{equation}
for less than half-filling. This relation expresses the equivalence
between the existence of isosbestic points and the formation of a
suppression of DC conductivity $\text{\ensuremath{\sigma}}_{DC}\left(\mu\right)$
resp. $\text{\ensuremath{\sigma}}_{DC}\left(n\right)$ around half-filling.

\section{\label{app:AppA}high-frequency limit}

\begin{figure}[t]
\centering{}\includegraphics[scale=0.65]{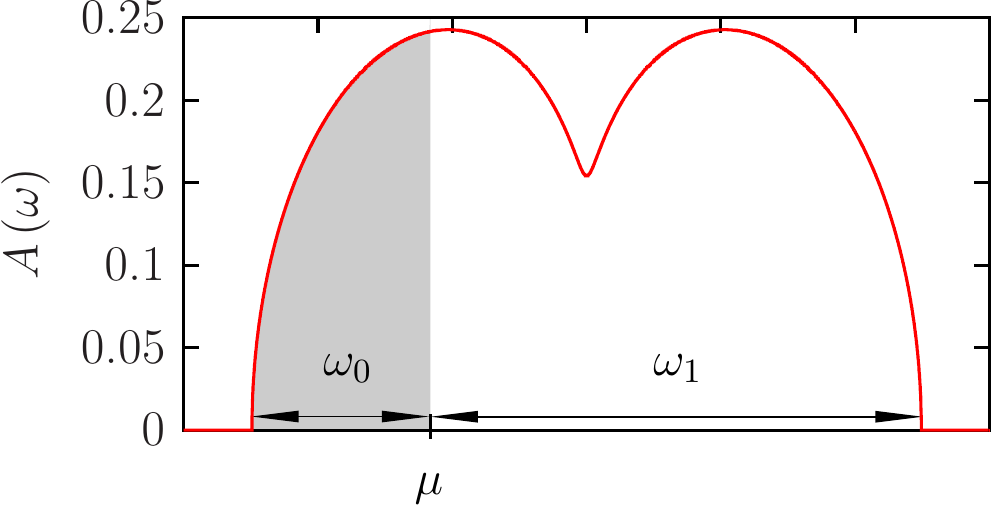} \caption{\label{fig:Fig9}DOS with finite support ($\mu<0$, the shaded area
indicates filled states): The term involving $A_{\epsilon}\left(-\omega\right)$
in Eq.~\eqref{eq:cond_deriv} vanishes for all frequencies $\omega>\omega_{0}$,
while $A_{\epsilon}\left(\omega\right)>0$ for all $\omega_{0}<\omega<\omega_{1}$.
Thus, the derivative $\partial\sigma\left(\omega_{0}\right)/\partial\mu$
will necessary be positive for all $\omega_{0}<\omega<\omega_{1}$;
for even higher frequencies it vanishes identically.}
\end{figure}

We determine the sign of $\partial\sigma(\omega_{0})/\partial\mu$
in the high-frequency limit, taking $\mu<0$. In this case, the Green
function and the self-energy are shifted to higher energies by $\mu$
compared to the half-filled case. For a symmetric DOS with finite
bandwidth (se Fig.~\ref{fig:Fig9}), it follows from Eq.~\eqref{eq:cond_deriv} that there exists
a frequency $\omega_{0}>0$ with $A(\omega_{0})>0$ and $A(-\omega_{0})=0$
(c.f. Fig.~\ref{fig:Fig3}). This yields $\partial\sigma(\omega_{0})/\partial\mu>0$
because the contribution involving $A_{\epsilon}(-\omega_{0})$ in
Eq.~\eqref{eq:cond_deriv} must identically vanish due to the finite
support of $\rho(\epsilon)$ and $\widetilde{\rho}\left(\epsilon\right)$.
Symmetric densities of states with infinite bandwidth behave similarly:
There exists a $\omega_{0}$ with $A\left(\omega\right)>A\left(-\omega\right)$
for all $\omega>\omega_{0}$ because of the stronger suppression of
the contribution involving $A\left(-\omega\right)$. This analogously
leads to $\partial\sigma(\omega_{0})/\partial\mu>0$.

\section{\label{sec:Shape-invariance}Shape invariance of single-particle
quantities with respect to doping}

Within DMFT, the self-energy of the model can be expressed explicitly
in terms of the interacting Green function {[}\cite{brandt1989tac,RevModPhys.75.1333}{]}
($w_{0}=1-w_{1}$):
\begin{equation}
\Sigma(\omega)=\frac{U}{2}-\frac{1}{2}\left[\frac{1}{{\displaystyle \mathrm{G}(\omega)}}\pm\,\sqrt{\left(U-\frac{1}{{\displaystyle \mathrm{G}(\omega)}}\right)^{2}+\frac{{\displaystyle 4w_{0}U}}{\mathrm{G}(\omega)}}\right],\label{eq:Sigma_of_G}
\end{equation}
 representing the exact summation of the skeleton expansion for $\Sigma(\omega)$.
In the low-$U$ regime an expansion of $\Sigma(\omega)$ in terms
of the small parameter $UG(\omega)$ gives ($w_{1}=1/2$)
\begin{equation}
\Sigma(\omega)=\frac{1}{2}U+\frac{1}{4}{\it U}^{2}G(\omega)-\frac{1}{16}{\it U}^{4}G^{3}(\omega)+\mathcal{O}[U^{6}].\label{eq:Sigma_of_G_expnd}
\end{equation}
 representing the first few terms of its skeleton expansion. As a
consequence of Eq.~\eqref{eq:imp_prob}, the model has an especially
simple doping dependence, where a change of $\mu$ (for constant $w_{1}$)
merely shifts the Green function and the self-energy, leaving their
shape invariant. Mathematically, this is expressed by the fact that
$\omega$ and $\mu$ enter $G(\omega)$ and $\Sigma(\omega)$ only
through $\omega+\mu$. To see that, we consider the inverse {[}\cite{RevModPhys.68.13}{]}
of the Hilbert transform $R\left[G(\omega)\right]$, expressing $\omega+\mu-\Sigma(\omega)$
exclusively through $G(\omega)$. Equation \eqref{eq:imp_prob} then
takes the form
\begin{eqnarray}
G(\omega) & = & \frac{1-w_{1}}{\omega+\mu-R\left[G(\omega)\right]+G^{-1}(\omega)}+\label{eq:G_through_G}\\
 &  & \frac{w_{1}}{\omega+\mu-R\left[G(\omega)\right]+G^{-1}(\omega)-U},\nonumber
\end{eqnarray}
 implicitly defining the Green function as function of $\omega+\mu$,
i.e., $G(\omega)=F\left(\omega+\mu\right)$. For the special case
of a semi-elliptic DOS, Eq.~\eqref{eq:G_through_G} leads to a cubic
equation, which can be solved explicitly {[}\cite{PhysRevB.45.2267}{]}
for $G\left(\omega+\mu\right)$.


\begin{thebibliography}{59}
\bibitem [{Lin()}]{Linear-scale}%
  \BibitemOpen
  \href@noop {} {}\bibinfo {note} {Of course, this width depends on the linear
  scale of the axes; only if the curves cross in one mathematically defined
  point is the linear scale irrelevant, see the discussion in Sec.
  \ref{sub:Sharpness-of-crossing} and Appendix \ref{sec:SharpnessS}}\BibitemShut {NoStop}%
\bibitem [{VIN()}]{VINV}%
  \BibitemOpen
  \href@noop {} {}\bibinfo {note} {In Refs.
  [\onlinecite{PhysRevB.82.064404}-\onlinecite{PhysRevLett.108.237204}] the
  crossing point observed in the specific heat curves $C(H,T)$, i.e., the near
  independence of $C(H,T)$ on the magnetic field $H$ at some temperature
  $T^{\star}$, is referred to as ``Vollhardt invariance''.}\BibitemShut {Stop}%
\bibitem [{Two()}]{Two-fluid}%
  \BibitemOpen
  \href@noop {} {}\bibinfo {note} {More generally, this applies in particular
  also to any kind of two-fluid model employed, for example, in
  phenomenological theories of superconductivity and superfluidity. There the
  density of the two components (e.g., the normal and superfluid component)
  depend on temperature while the total density is constant:
  $n=n_{1}(T)+n_{2}(T)=\text{const}$. Properties of the system are then
  described by the superposition of the two components, leading to a special
  dependence of quantities $f(T,X)$ on $T$ and $X$ of the form
  $f(T,X)=n_{1}(T)f_{1}(X)+[n-n_{1}(T)]f_{2}(X)$. This implies the crossing of
  curves for different temperatures $T$ at a single point $X^*$ determined by
  $f_{1}(X^*)=f_{2}(X^*)$.}\BibitemShut {Stop}%
\bibitem [{Com()}]{CommentCuprate}%
  \BibitemOpen
  \href@noop {} {}\bibinfo {note} {Other combinations (e.g. $T_{1}=40K$ and
  $T_{2}=80K$ etc.) showed no significantly different behavior of
  $\widetilde{\chi}\left(\omega,T\right)$ and $\chi_{2}(\omega)$, apart from an
  eventual increase in noise in $\chi_{2}(\omega)$, which can become larger
  than the numerator in the expression for $\chi_{2}(\omega)$ for combinations
  like $T_{1}=14K$ and $T_{2}=20K$.}\BibitemShut {Stop}%
\end{thebibliography}
\end{document}